# THE UMBRAL-PENUMBRAL BOUNDARY IN SUNSPOTS IN THE CONTEXT OF MAGNETO-CONVECTION

D. J. Mullan & J. MacDonald

Department of Physics and Astronomy, University of Delaware, Newark DE

ABSTRACT

Jurcak et al (2018) have reported that, in a sample of more than 100 umbral cores in sunspots, the umbral-penumbral boundary (UPB) is characterized by a remarkably narrowly-defined numerical value (1867 G) of the *vertical* component of the magnetic field. Gough and Tayler (1966), in their study of magneto-convection, showed that the onset of convection in the presence of a magnetic field is controlled by a parameter $\delta$ which also depends on the *vertical* component of the field. Combining the Jurcak et al result with various empirical models of sunspots leads us to propose the following hypothesis: the UPB occurs where the vertical field is strong enough to increase the effective adiabatic temperature gradient by ≥100% over its non-magnetic value.

Keywords: Sun-magnetic field, Sun: sunspots; Sunspots: umbra; Sunspots: penumbra

## 1. INTRODUCTION

In 2011, Jurcak (2011) reported on a study of magnetic field properties at a specific location in a small sample of sunspots. The specific location to which Jurcak (2011) paid attention was the umbral-penumbral boundary (UPB). In that paper, he commented that, to his knowledge, "no one [had] yet tried to estimate the properties of the magnetic field *right at the penumbra boundaries*" (our emphasis added). The boundary which is of primary interest in the present paper is the one where the penumbra is in contact with the umbra, i.e. the UPB. (The other boundary, between penumbra and photosphere, is not part of our discussion.) Jurcak's goal in 2011 was to observe the magnetic parameters at the UPB and to "find out whether they are the same for sunspots of different sizes, and if they are even constant along the boundaries in a given sunspot".

In a subsequent extended study of 79 different active regions, Jurcak et al (2018) reported on their analysis of full Stokes profiles of an Fe I line obtained by the Hinode satellite between 2006 and 2015 for spots in which the umbral areas were ≥10 Mm$^2$. They discovered that at the UPB, "the vertical component of the magnetic field strength [$B_v$] does not depend on the umbra size, or on its morphology, or on the phase of the solar cycle". They found that the numerical value of $B_v$ at the UPB has a most probable value of 1867 G, with a 99% likelihood of lying in the range 1849-1895 G.

This is a remarkable discovery. Jurcak et al. noted that "it gives fundamental new insights into the magneto-convective modes of energy transport in sunspots".

Support for the discovery of Jurcak et al (which was derived on the basis of many different active regions) has been provided by Schmassmann et al (2018) who followed a single stable spot as it crossed the disk. They found that, in the course of 10 days of observing, the vertical component $B_v$ of the magnetic field at the UPB remained constant with a r.m.s. deviation of less than 1%. To be sure, Schmassmann et al found that the numerical value of $B_v$(UPB) was 1693 G, which is discrepant from the value reported by Jurcak et al by "some 175 G". However, Jurcak et al used the Hinode SP instrument for their work, while Schmassmann et al used SDO/HMI. The two studies relied on different spectral lines, different spectral resolutions, different stray light corrections, etc. In view of this, Schmassmann et al attribute the discrepancy between $B_v$(UPB) = 1867 G (Jurcak et al) and $B_v$(UPB) = 1693 G (Schmassmann et al) to "differences in the experimental setup and analysis methods". Our goal here is to point out a connection between this discovery and one particular model of magnetoconvection.

## 2. THE GOUGH-TAYLER CRITERION FOR ONSET OF MAGNETOCONVECTION

Gough and Tayler (1966: hereafter GT) derived a criterion for the onset of convective instability in an electrically conducting gas which is permeated by a magnetic field. In order to set the stage for a discussion of GT, we first consider the case of a compressible medium which does not contain any magnetic field.

### 2.1. Onset of convection in a non-magnetic medium.

In a medium which does not contain magnetic fields, the well-known Schwarzschild criterion is valid: convection sets in when the temperature gradient is steeper than the adiabatic gradient. Expressing the gradients in logarithmic terms, where $\nabla \equiv d \ln T / d \ln p$ is the local temperature gradient with respect to gas pressure $p$, the Schwarzschild criterion is $\nabla > \nabla_{ad}$. In a gas which is non-ionizing, $\nabla_{ad}$ can be written as $(\gamma - 1)/\gamma$ where $\gamma$ is the adiabatic exponent (e.g. Mullan 2009, eq. 6-13). In a monatomic gas, $\gamma = 5/3$, and therefore $\nabla_{ad} = 0.4$.

How permissible is it for us to assume that the double conditions of monatomic and non-ionizing are applicable to the gas in the photosphere of a sunspot? To answer this, we first consider the conditions in the non-magnetized portions of the quiet Sun.

In the quiet Sun, the major constituents (H, He) are only weakly ionized: at $T$ = 6000 K, the fraction of ionized H is of order 1 part in 20,000 (e.g. Mullan 2009, p. 59), and He is even less ionized. In the umbra of a sunspot, where the effective temperature is lower than photospheric, about 4160 K (Bray & Loughhead 1964, p. 107), the degrees of ionization of H and He are even smaller. The only elements which will be ionized in a sunspot photosphere will be elements with the lowest ionization potentials, such as the alkali metals. These have such small abundances in the Sun that we will make no significant error if we proceed as follows: the criterion of non-ionizing gas is readily applicable to gas in the umbral photosphere.

In what follows, we shall require that the gas in a sunspot be capable of being "interfered with" by magnetic fields. To ensure that such coupling can occur at all, there must be *some* finite value for the electrical conductivity. That is, the gas in the sunspot cannot be absolutely neutral in an electrical sense: the gas must be at least partially ionized. However, when we consider in detail the physical processes which occur when magnetic fields interfere with convective flow patterns, we shall find that even in the presence of the small amount of partial ionization which exists in the umbra of a sunspot, the interaction between field and gas can be modeled with high confidence by assuming that the gas is infinitely conducting. (For quantitative details in support of this claim, the reader is referred to the Appendix.) In view of this, we shall assume explicitly that the electrical conductivity is infinite in the calculations to be reported below (in Section 2.2).

What about the requirement of "monatomic"? This assumption could be suspect if the temperature in the umbra were to be low enough for abundant molecules to form. To address this, we note that Vardya (1966) has analyzed the equilibrium abundances of more than 100 molecular species, atoms, as well as positive and negative ions, in the atmospheres of K and M dwarfs: these stars have effective temperatures ranging from 4410 K for K5 stars to 3920 K for M0 stars to 2660 K for M8 stars. The umbral effective temperature mentioned above (4160 K) falls between the temperatures of a K5 and an M0 dwarf in Vardya's list. Therefore, if we examine the molecular abundances in an M0 dwarf, we can get an impression of what to expect as upper limits on molecular abundances in the (slighter hotter) umbra of a sunspot. Vardya finds that in an M0 star, the most abundant constituent in the atmosphere is monatomic hydrogen. A molecular species ($H_2$) does not become the dominant constituent until we get to stars as cool as M2, with effective temperatures of only 3500 K. Therefore, in the umbra of a sunspot, Vardya's results suggest that we are safe in assuming that the gas is effectively monatomic. This conclusion helps to strengthen the "non-ionizing" condition mentioned in the preceding paragraph: if molecules were to be present in abundance in the gas in the umbral photosphere, we would have to incorporate the effects of dissociation in the same way as those of ionization when estimating the value of the adiabatic exponent $\gamma$.

In view of these considerations, we expect that we will not make any significant error if we write the Schwarzschild condition for the onset of non-magnetic convection in the gas which exists in a sunspot *umbra* in the following form: $\nabla > 0.4$. The numerical value of 0.4 on the r.h.s. of this inequality will be important in what follows.

2.2. Onset of convection in a medium with a magnetic field

Now we turn to the case of a medium in which a magnetic field is present, such as GT considered. In such a medium, if the electrical conductivity is infinitely high, the field and the gas become "frozen together" such that any attempt to force the gas to move in some direction (e.g. by participating in the overturning motions associated with convection) inevitably leads to a forcing of the field to move as well. In response to any imposed force (e.g. buoyancy), not only must the inertia of the gas (with its finite energy density) be taken into account: the energy density of the magnetic field will also contribute to how the medium will react to the imposed force. As a result, the onset of convection is likely to be impeded in some way by the presence of the field. No longer does the Schwarzschild criterion suffice to determine the onset of convection.

In order to quantify the criterion for the onset of convection instability in a perfectly conducting gas in the presence of a magnetic field, GT relied on an energy principle which was originally developed by Bernstein et al. (1958) in the context of laboratory plasmas. The approach is as follows: starting with an initial configuration of magnetic field and gas, a small perturbation is applied and the change ΔW in the total energy of the system is computed. If it can be shown that, for all permissible small perturbations, ΔW is a positive quantity, then the configuration can be regarded as stable. But if there exists even one example of permissible perturbations which leads to a reduction in ΔW, then the configuration is unstable.

GT found that a condition which would ensure magneto-convective stability could be written in the form

$$\frac{B_v^2}{B_v^2 + 4\pi\gamma p} > \nabla - \nabla_{ad}. \tag{1}$$

Here, $\gamma$, $p$, $\nabla$ and $\nabla_{ad}$ have the same meanings as above. (Note that we have adjusted eq. (1.2) of GT by including a factor of $4\pi$ in the denominator: the reason for this is that GT used rationalized Gaussian units whereas we use Gaussian c.g.s. units.)

We draw special attention to a quantity which did not appear at all in the Schwarzschild criterion, but which appears in the GT criterion: $B_v$. This is *not* the total magnetic field strength: instead, it represents only one of the components of the vector magnetic field, namely the *vertical* component of the field. Using the above formula, we can re-write the GT result in terms of a criterion for the onset of convective *in*stability in the presence of a magnetic field as follows:

$$\nabla > \nabla_{ad} + \delta \tag{2}$$

where $\delta = B_v^2 / (B_v^2 + 4\pi\gamma p)$. In contrast to the Schwarzschild criterion, which stated that convection would set in as soon as $\nabla$ grows to a value which exceeds $\nabla_{ad}$, the GT criterion states that, in the presence of a (vertical) magnetic field, convection will not set in until $\nabla$ exceeds the larger numerical value $\nabla_{ad} + \delta$. Note that the larger the value of $\delta$, the larger must $\nabla$ become in order for convection to set in, i.e. the steeper must the temperature gradient become before convection can occur. Thus, the larger $\delta$ is, the greater is the effect of the magnetic field in inhibiting the onset of convection. In this sense, $\delta$ can be regarded as a magnetic inhibition parameter.

The principal point of the present paper is that the component of the magnetic field which appears in the GT criterion, i.e. $B_v$, is the same component that Jurcak et al. have identified as playing a fundamental role at the umbral-penumbral boundary in sunspots. This leads us to consider that it might be profitable to regard the UPB as the site where local conditions ensure that the onset of convection is required to satisfy not the Schwarzschild criterion, but rather the more difficult criterion described by GT.

On a practical note, no real star contains material with infinite conductivity. Therefore, we need to ask: to what extent can we apply the GT criterion to a medium where the conductivity is finite? This issue is addressed in an Appendix below. The conclusion is that in the context of convective flows in the kinds of stars in which we are interested, the presence of finite conductivity does not have any significant effect on our conclusions.

### 2.3. Numerical considerations

Recalling the discussion in Section 2.1, it is worthwhile to write the GT criterion for magneto-convective onset as $\nabla > 0.4 + \delta$. In this form, we see that if it can be shown that there are astrophysical cases where $\delta$ is *small* compared to 0.4, we expect that such cases should have convective properties that are only slightly different from those of non-magnetic convection. But if, on the other hand, we can identify cases in which $\delta$ approaches, or even exceeds, a numerical value of 0.4, then we expect the convective properties in such cases should deviate significantly from those of non-magnetic convection. In the next Section, we turn to examples in which the value of $\delta$ has been found to be small compared to 0.4. In Section 4, we shall turn to the opposite limit, when $\delta$ can definitely *not* be considered to be small compared to 0.4.

## 3. MAGNETO-CONVECTIVE MODELLING EFFORTS IN STARS: "SMALL" CHANGES IN THE THRESHOLD FOR CONVECTIVE ONSET

In 2000, Leggett et al reported on measurements of infra-red fluxes from cool dwarfs which allowed bolometric luminosities to be determined with higher precision then before. For the first time, the numerical values of stellar radii could then be obtained for a sample of several dozen M dwarfs with errors of no more than 10-15%. When the data were compared with stellar models, these error bars were good enough to suggest the following conclusion: "active M dwarfs have radii which are systematically too large [compared to models] for their effective temperatures" (Mullan & MacDonald 2001: hereafter MM01). Since active M dwarfs are known to be magnetic, the anomalously large radii led MM01 to explore the possibility that magnetism might alter the onset of convection sufficiently to cause global structural changes to stellar models. With that in mind, MM01 calculated stellar models in which the GT criterion was applied to the onset of magneto-convection. The resulting models, though exploratory in nature, were indeed found to have larger radii (for a given stellar mass) than non-magnetic models would predict.

The greatest uncertainty in applying the GT criterion to a star in 2001 was (and still is) our lack of information about the radial profile of the inhibition parameter $\delta$. The place where it is easiest to evaluate $\delta$ is in the photosphere of a star, where gas pressure and surface field strength can in principle be measured. But how are we to proceed at greater depths below the surface? Following Ventura et al (1998), the simplest approach would be, once the surface value of $\delta$ has been decided upon, to set $\delta$ equal to the same constant value at all radii. Other profiles of $\delta(r)$ can also be explored, but MM01 found that the overall results did not differ greatly between the various choices for the $\delta(r)$ profile. Models of stars with masses ranging from 0.375 $M_\odot$ down to 0.1 $M_\odot$ were explored in which $\delta$ was assigned values ranging from 0.005 to 0.07. Those ranges of $\delta$ were selected with a view in mind (suppression of convection in the core) which has since been recognized as inappropriate for cool dwarfs: the required magnetic fields would be much too strong to be generated by stellar dynamos (e.g. MacDonald & Mullan 2012: MM12). This realization led MM12 to compute a model which, abandoning the $\delta(r)$ = constant profile, instead imposed a "ceiling" value of $10^6$ G on the field strength. Such a ceiling ensures that the value of $\delta(r) \to 0$ as we approach the center of the star. Subsequently, the MM12 choice of "ceiling" field was shown (Browning et al. 2016) to be the strongest field that could plausibly survive a number of instabilities in a low-mass star in the course of evolutionary times.

The goal of our magneto-convective models has been to replicate observed radii and luminosities in low-mass stars with known ages. In the presence of a "ceiling" on the field in the deep interior, successful fitting of empirical radii requires us to assign increasing values of $\delta$ at the surface of the star as the value of the ceiling field decreases. As a result, the largest values of $\delta$ which have been found to be necessary to replicate the empirical stellar radii and luminosities have emerged from models in which the "ceiling" field was limited to a very low value. What might the lowest value of the "ceiling" field be in stars? Various 3-D modeling efforts in dynamo field generation suggest that low mass stars can readily generate fields of 10-20 kG: see MacDonald & Mullan (2017: MM17) for a summary of those dynamo models. In view of the dynamo results, MM17 selected 10 kG as the "ceiling" field, and then obtained models to fit the empirical data on a sample of 14 stars with well-defined ages. MM17 found values of $\delta$ as follows. The two lowest mass stars, with masses of 0.22 $M_\odot$, required $\delta$ = 0.013-0.051 and 0.038±0.015. The two highest mass stars ($M$ = 0.852, 0.862 $M_\odot$) required $\delta$ = 0.03-0.05. The smallest values of $\delta$ (0.018-0.033) were found in a star with mass 0.23 $M_\odot$, while the largest values of $\delta$ occurred in a fast-rotating binary (CM Dra) in which the components were required to lie in the rather wide range $\delta = 0.03$-$0.11$.

Among the MM17 sample of 14 stars with well-defined ages, the mean value of $\delta$ determined by MM17 ranges from 0.010 to 0.095, with a median value of 0.043. In view of the fact that these results were obtained with a ceiling field of only 10 kG (likely to be actually weaker than the fields which exist inside a low-mass star), the $\delta$ values described above should be regarded as upper limits: if we were to allow the "ceiling" field to be stronger than 10 kG, then we would expect to find even smaller values of $\delta$ in the best-fit solutions.

In summary, the stellar models described in this section are found to provide fits to the empirical radii and luminosities using values of $\delta$ which have median values of 0.043 or smaller.

Should this result be considered as a "large" value of $\delta$, or as a "small" value of $\delta$? To answer this, we must compare the value of $\delta$ with the threshold $\nabla = \nabla_{ad} = 0.4$ for the onset of non-magnetic convection. We see that, in the stars which have been modelled by MM17, convection sets in when the temperature gradient is larger than the non-magnetic threshold by an amount which is on average no more than 10%. In this sense, the magneto-convective solutions obtained in MM17 can be regarded as relatively small (typically <10%) perturbations on the solutions which would be obtained in the non-magnetic limit. The smallness of the changes relative to non-magnetic models can be appreciated from the differences between the stellar radii which they predict and the radii predicted by non-magnetic models. These differences amounted to 10-15% (with large error bars) for the earliest data (Leggett et al. 2000), but in subsequent data, the changes were found to be only a few percent. From a historical perspective, it was not until the precision of the empirical determinations of the masses and radii became as good as a few percent that computation of magneto-convective models really became worth the effort. As Torres et al. (2010) have stated: "Only data with errors [in the mass] below ∼1–3% provide sufficiently strong constraints that models with inadequate physics can be rejected".

In the context of the discussion on Section 2.3 above, we expect that, as long as $\delta$ has numerical values which are no more than 10% of $\nabla_{ad}$, then the changes which will be produced in the observable physical quantities such as luminosity and radius (relative to non-magnetic solutions) will remain "small", i.e. 10% or less.

As a caveat in the above discussion, we recognize that although the numerical value $\nabla_{ad} = 0.4$ is valid for the objects of primary concern in this paper (i.e. the umbrae of sunspots, where gas temperatures are of order 4000 K), this is not necessarily true for some of the objects which have been subjected to magneto-convective modelling by MM17. In MM17, all but one of the target stars have spectral types which are M2 or later. According to Vardya (1966), in such stars, $H_2$ molecules may be the dominant constituent of the atmosphere. In the coolest stars ($T < 3000$ K, i.e. too cool for $H_2$ dissociation), the availability of rotational degrees of freedom will reduce $\gamma$ from 5/3 towards a value of 7/5, leading to $\nabla_{ad} \approx 0.3$. In stars which are hot enough to dissociate $H_2$, the extra degrees of freedom will reduce $\gamma$ further, leading to $\nabla_{ad}$ even smaller than 0.3. How small might $\nabla_{ad}$ become in such environments? Only a detailed model would provide a reliable answer: however, if we examine an analogous case (i.e. ionization of H atoms) in a model of the solar envelope which lists the relevant information (Baker & Temesvary 1966), we find that $\nabla_{ad}$ has a minimum value of 0.12. If this were to be a reliable value of the minimum $\nabla_{ad}$ in the MM17 stars, then our median value of $\delta = 0.043$ would require that for convection to set in, the temperature gradient would have to be 35% larger than in the non-magnetic case. This could probably not be classified formally as a "small" perturbation. But a factor of 35% still lies well below the case which occurs in the umbra of a sunspot: in the latter case, we shall find (Section 4) that in order for convection to set in in the presence of the fields which exist at the UPB, the temperature gradient must exceed the non-magnetic gradient by a factor of 100% or more.

## 4. MAGNETO-CONVECTION IN SUNSPOTS: "LARGE" CHANGES IN THE THRESHOLD FOR CONVECTIVE ONSET

The work of Jurcak et al. (2018), with its well-defined value of $B_v$ = 1867 G at the UPB, suggests that it might be informative to consider this field in the context of the magnetic inhibition parameter $\delta$. To do this, we need to know the gas pressure $p$ at some reference level: for the sake of definiteness, we choose the reference level at the location where the continuum optical depth $\tau$ has a value of unity. An anonymous referee has pointed out that Jurcak et al (2018) undertook their measurements of $B_v$(UPB) using the FeI 6302Å line which corresponds in a continuum optical depth $\tau$ lying between 0.1 and 0.01. As a result, strictly speaking, the magnetic information provided by the FeI line does not refer to the same level in the atmosphere as the pressures (at $\tau$ = 1) given in Table 1. For example, referring to the models of Maltby et al (1986), the gas pressure at $\tau$ = 0.1 is lower by a factor of order 3 compared to the pressure at $\tau$ = 1. In principle, we anticipate that if we were to use the (smaller) gas pressures at the level in the atmosphere to which $B_v$(UPB) actually refers, i.e. $\tau \approx 0.1$, then the numerical value of the magnetic inhibition parameter $\delta \sim 1/p$ would become *large*r than the values listed in Table 1, perhaps by as much as a factor of 3.

In a survey of the literature, we have identified 13 sunspot models which provide us with numerical values of $p(\tau=1)$. For each model, we have combined the $p(\tau=1)$ value with the Jurcak et al. (2018) value of $B_v$ = 1867 G to obtain a value for $\delta(\tau=1)$. Results are listed in Table 1. (With regard to the sunspot models, we recognize that inside an umbra, the magnetic field strength may well vary as we move from radial locations at the center of the umbra to radial locations close to the UPB: e.g. Broxon 1942. These variations in field strength could be accompanied by gas pressure variations as we move from umbral center to UPB. We assume that the models listed in Table 1 are providing gas pressures which are in some sense a physically meaningful average value which is representative of the conditions in the gas at $\tau$=1.)

The models in Table 1 were derived by a variety of techniques. Some used observations of lines, some used the continuum. The models based on lines used a curve of growth technique in the earliest models, but switched to inversion of Stokes parameters data in more recent work. The models which were derived from continuum data span a range of wavelengths which is broad enough to include the minimum in H-minus absorption (at 1.6μm). In general, the (7) continuum models are expected to probe conditions relatively deep in the spot, whereas the (6) line-based models would have probed conditions somewhat higher in the atmosphere.

TABLE 1. Models of sunspot umbrae

| Reference for model | $p(\tau=1)$ (dyn/cm$^2$) | $\delta(\tau=1)$ | Notes |
|---|---|---|---|
| Michard (1953) | 3.55 x 10$^4$ | 0.824 | 0.3-2.3µm contin. |
| Mattig (1958) | 2.63 x 10$^5$ | 0.388 | Curve of growth |
| Fricke & Elsasser (1965) | 6.31 x 10$^4$ | 0.725 | Curve of growth |
| Yun (1971) | 2.82 x 10$^5$ | 0.371 | Contin. |
| Moe & Maltby (1974) Model B | 3.02 x 10$^5$ | 0.355 | 0.4-1.7µm contin. |
| Moe & Maltby (1974) Model D | 3.98 x 10$^5$ | 0.295 | " |
| Maltby et al. (1986) Model L | 2.37 x 10$^5$ | 0.413 | 0.5-2.5µm contin. |
| Maltby et al. (1986) Model E | 3.06 x 10$^5$ | 0.352 | " |
| Maltby et al. (1986) Model M | 2.68 x 10$^5$ | 0.383 | " |
| Collados et al. (1994) warm | 1.85 x 10$^5$ | 0.474 | FeI line profiles |
| Collados et al. (1994) cool | 3.06 x 10$^5$ | 0.352 | " |
| Socas-Navarro (2007) Model A | 1.74 x 10$^5$ | 0.489 | CaII+FeI line profiles |
| Socas-Navarro (2007) Model B | 1.59 x 10$^5$ | 0.511 | " |

Of course, the investigators who obtained the models listed in Table 1 were in no cases aware of the result of Jurcak et al. (2018) regarding the existence of a unique value of $B_v$ at the UPB. Therefore, although the results of GT were already in the literature when 10 of the above models were being developed, it would have been unlikely that a calculation of the GT inhibition parameter $\delta$ would have been undertaken.

But now, with access to information about the very component of the field which enters into the GT formula for $\delta$, the models can be used to evaluate $\delta(\tau=1)$ in each case. When we average the values of $\delta(\tau=1)$ in table 1 for the continuum-based models, we find $<\delta(\tau=1)> = 0.43$. Repeating the calculation for the line-based models, we find $<\delta(\tau=1)> = 0.49$. Averaging all 13 models, we find $<\delta(\tau=1)> = 0.46$. And if we are to include the 3-fold correction mentioned above to allow for the reduced gas pressure at $\tau= 0.1$, we would find $<\delta> \approx 1.38$.

In the context of the discussion on Section 2.3 above, we now revisit the question: are the values of $\delta$ to be considered "small" or "large"? Once again, it is necessary to compare the $\delta$ values with the critical value ($\nabla_{ad}$) of the adiabatic temperature gradient in a non-magnetic medium. Whereas in global stellar models, we found that the value of $\delta$ was small (<10%) compared to the critical $\nabla_{ad} = 0.4$, this is no longer true in the case of the UPB in a sunspot. The results of Jurcak et al (2018), in combination with eq. (2) above, make it clear that the temperature gradient required for convection to set in at the UPB is

$$\nabla > \nabla_{ad} + \delta = 0.4 + 0.46. \qquad (3)$$

Therefore, the sunspot models in Table 1 indicate that the onset of convection at the UPB requires the temperature gradient to exceed the adiabatic gradient by a factor which is by no means "small". Instead, as is obvious from eq. (3), the superadiabaticity (i.e. the excess of the temperature gradient above $\nabla_{ad}$) at the UPB must be at least 100%. And if we were to include formally the effects of ionization which occur even in sunspots among some of the low-abundance "metals", the value of $\nabla_{ad}$ would be reduced somewhat below 0.4. In that case, our "GT correction" of 0.46 would represent an increase in the requisite temperature gradient that could be well in excess of 100%. And if we were to allow for the reduction in gas pressure between the levels in the atmosphere where $\tau = 1$ and $\tau = 0.1$ (see the first paragraph at the start of Section 4), such a reduction in pressure would lead to a superadiabaticity (i.e. a value of $\delta$) which could be as large as 1.38 in eq. (3). This would lead to the conclusion that the excess of the temperature gradient above $\nabla_{ad}$ at the UPB must be well in excess of 100%.

Such gross departures from the non-magnetic criterion for convective onset in an umbra suggest that gross departures from the non-magnetic photon flux should arise. In fact, the empirical effective temperature of an umbra is in one case (Bray & Loughhead 1964; p. 114) listed as 4480 K. Comparing this with the effective temperature of the quiet Sun (5740 K), we find that the bolometric flux emerging from the quiet Sun is greater than that from the umbra by a factor of 2.7. That is, the quiet Sun emits 170% more flux than the umbra does. Clearly, with an amplitude of 170% for the difference, we are not dealing here with "small perturbations" to the energy flux. The observational effects which arise from the presence of the magnetic field in sunspots are quite different from the "small perturbations" which have been observed in the equivalent physical parameters in stars (as described in Section 3).

We note that, in the GT model, the approach to convective transfer is essentially one-dimensional, such as occurs when we model a spherically symmetric star. However, shortly after the paper by Jurcak et al (2018) appeared, three-dimensional models of convection in stars of various spectral types were reported by Salhab et al (2018), for both magnetic and non-magnetic conditions. The results which are presented in Figure 10 of Salhab et al are of particular interest in the context of the present paper: they show numerical values for the superadiabaticity as a function of optical depth. For a solar model, Salhab et al find that the maximum value of superadiabaticity is about 1.3: therefore, the value of 1.38 mentioned above for our evaluation of the quantity $\delta$ at the UPB does not appear at all inconsistent with the maximum superadiabaticity which has been found in 3-D radiative models of the Sun.

## 5. CONCLUSION

The discovery by Jurcak et al. (2018) that the umbra-penumbral boundary in a sample of order 100 sunspots is defined by a narrowly-constrained value of 1867 ± 18 G for $B_v$, the *vertical* component of the field, is remarkable. There is no indication that other components of the field, or the total field strength, are limited to such narrow windows. Why should the vertical component of the field be the only component to be constrained to lie within such a narrow window?

In this paper, we suggest that a possible reason for this behavior can be found in one particular version of the criterion for the onset of convection in the presence of a magnetic field. Gough and Tayler (1966: GT) derived such a criterion and found that convection will set in only when the (logarithmic) temperature gradient $\nabla$ exceeds a limit which is no longer equal to the simple Schwarzschild value ($\nabla_{ad}$). Instead, the GT criterion for onset of convection is found to be $\nabla > \nabla_{ad} + \delta$. In this new expression, $\delta$ is a positive definite quantity which depends on two physical parameters: the gas pressure, and the *vertical* component of the magnetic field.

We suggest that the appearance of the vertical component of the field strength as an essential term in the GT criterion can explain why Jurcak et al. (2018) have identified an essentially unique value for $B_v$ at the location where the pronounced dimming associated with the umbra occurs.

Quantitatively, the values of $\delta$ which have been derived from fitting global physical parameters of low-mass *stars* (See Section 3) are found to be no more than a few percent of $\nabla_{ad}$. With such small values, the corresponding magnetic fields do not alter greatly the Schwarzschild criterion for the onset of convection. As a result, magnetic effects give rise to only relatively minor perturbations (a few percent) to the radii and luminosities of low-mass stars. In fact, non-standard physics (such as magnetic effects) could not even begin to be identified confidently in low-mass stars until the measurements of masses and radii had improved to the point where the errors were reduced to no more than a few percent (Torres et al. 2010).

On the other hand, now that Jurcak et al. have provided reliable measurements of $B_v$ at the umbral-penumbral boundary, we can establish that the values of $\delta$ at the UBP are not at all small relative to $\nabla_{ad}$. Quite the contrary: at the $\tau=1$ level in an umbra, we find that the value of $\delta$ is of order 100% or more of $\nabla_{ad}$. Therefore, if convection is to set in in such conditions, it is not sufficient for $\nabla$ merely to exceed $\nabla_{ad}$: instead, $\nabla$ is now forced to exceed a value which is the sum of $\nabla_{ad}$ plus another term which is at least as large as 100% of $\nabla_{ad}$. Thus, onset of convection in this case requires conditions which are grossly different from the non-magnetic case. In such conditions, it would be unreasonable to expect that only small (few percent) variations should occur in the luminosity. On the contrary, variations in energy flux of order 100% are expected to occur. We suggest that these large variations contribute to the significant dimming of a sunspot umbra relative to the photosphere.

## APPENDIX

**EFFECTS OF FINITE CONDUCTIVITY**

The GT approach relies formally on the assumption that the gas in question is infinitely conducting. However, in a real star, the gas does not have infinite conductivity. At first sight, this suggests that we could be in error if we were to apply the GT model to a real star. But when we examine the conditions in a real star quantitatively, we find that this is not a major difficulty in the context of the physics of convection. In fact, we claim that a convecting medium with finite conductivity can behave in a way which differs so slightly from the behavior of a medium with infinite conductivity that the error we would make turns out to be small (of order 1%).

To justify this claim, we first note that, in the presence of finite conductivity, the magnetic field and the gas are formally no longer "frozen together": instead, the field can drift relative to the parcel of gas in which it was contained at time $t=0$. This leads to a finite spatial separation $L_{ss}$ of field from its initial parcel of gas in the course of a certain time. We now need to ask: how much spatial separation $L_{ss}$ can occur during a time interval $T_c$ which has some relevance for the process of thermal convection? The answer to this question depends on a characteristic exponential decay time-scale $t_d$ for magnetic fields in a medium where the electrical resistivity $\eta$ is non-zero (Spitzer 1962, eq. 2-38): $t_d \approx 4\pi L_{ss}^2/\eta$ if $\eta$ is expressed in electromagnetic units. If the electrical conductivity $\sigma = 1/\eta$ is expressed in electrostatic units (esu), we find $L_{ss} = c\sqrt{(T_c/4\pi\sigma)}$, where $c$ is the speed of light.

The question now is: how do typical values of $L_{ss}$ in the Sun compare with a characteristic length scale $L_c$ of the convective flows which exist in the Sun's photospheric gas?

To evaluate the quantities $T_c$ and $L_c$ we turn to observations. Since convection in a star is highly turbulent, the convective flows occurs in individual eddies ("granules") which survive only for a finite time: in the Sun, this time is observed to be of order 5-10 minutes (Title et al. 1989). After that time, the eddy loses its identity, dissolves back into the turbulent medium, and eventually becomes part of a new eddy. Thus, a relevant time-scale $T_c$ for convection is 300-600 seconds. The electrical conductivity in the partially ionized gas which exists in the photospheric region of a sunspot umbra has been calculated (Bray & Loughhead 1964, p. 125) to be $\sigma = 10^{11}$ esu. Inserting $c = 3 \times 10^{10}$ cm s$^{-1}$, we can now evaluate the quantity $L_{ss}$ in the umbral photosphere: we find $L_{ss} \approx 5$-7 km. That is, in the course of one granule lifetime, the magnetic field can spatially separate from its original gas parcel to an extent of less than 10 km.

How does this spatial separation compare with the size of a convection cell? From their observations of granule sizes in the Sun, Title et al. (1989) report that "it is fair to say that that there is a characteristic granule [angular] size in the vicinity of 1.2-1.4 arc-seconds". This angular scale corresponds to a linear size of $L_c \approx 900$-1000 km. These are the horizontal dimensions which are typical of granules in the solar photosphere. Compared to these dimensions, the spatial separation of field and gas during a granule lifetime amounts to less than 1%.

In view of this small percentage, we see that *as far as magnetic interference with granule flows is concerned,* the gas in the photosphere of a sunspot behaves in essentially the same way as if it had infinite conductivity. Thus, the approach used by Gough & Tayler (1966) for quantifying the onset of convection in a magnetic field can be applied without significant error to the photospheric gas in a sunspot umbra.